# Regenerating Codes for Errors and Erasures in Distributed Storage

K. V. Rashmi, Nihar B. Shah, Kannan Ramchandran, *Fellow, IEEE*, and P. Vijay Kumar, *Fellow, IEEE*

*Abstract*—Regenerating codes are a class of codes proposed for providing reliability of data and efficient repair of failed nodes in distributed storage systems. In this paper, we address the fundamental problem of handling errors and erasures during the data-reconstruction and node-repair operations. We provide explicit regenerating codes that are resilient to errors and erasures, and show that these codes are optimal with respect to storage and bandwidth requirements. As a special case, we also establish the capacity of a class of distributed storage systems in the presence of malicious adversaries. While our code constructions are based on previously constructed Product-Matrix codes, we also provide necessary and sufficient conditions for introducing resilience in any regenerating code.

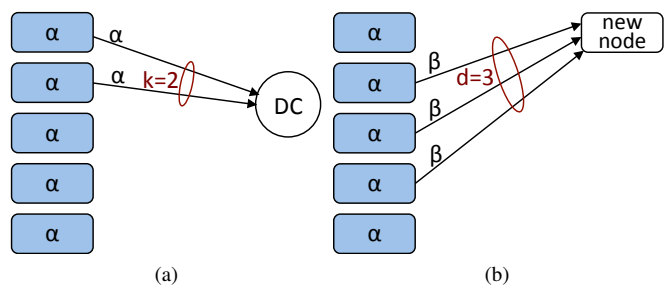

Fig. 1: An example of the system parameters under a regenerating code (in the absence of errors/erasures). The system comprises of $n = 5$ storage nodes: (a) reconstruction is accomplished from any $k = 2$ nodes, (b) repair from any $d = 3$ nodes.

## I. INTRODUCTION

Distributed storage systems play a vital role in today's age of big data. For cost considerations, these storage systems often employ commodity hardware, which makes failures a norm rather than an exception. In order to safeguard the precious data against such failures, the data is typically stored in a redundant manner. In this paper, we consider a distributed storage system consisting of $n$ storage nodes in a network, each having a capacity to store $\alpha$ symbols over a finite field $\mathbb{F}_q$. Data comprising $B$ symbols (the *message*) is to be stored across these $n$ nodes. An end-user (called a *data collector*) must be able to *reconstruct* the entire message by downloading the data stored in *any* $k$ of these $n$ nodes. It follows that such a system can tolerate failure of any $(n-k)$ nodes, and under solely this requirement, can be realised using any $[n, k]$ maximum distance separable (MDS) code.

Frequent node failures also call for efficient handling of the failure events. When a storage node fails, it is replaced by a new, empty node. This replacement node is required to obtain the data that was stored previously in the failed node, by downloading data from the remaining nodes in the network. We will term this process as *repair* or *regeneration* of a node. A typical means of accomplishing this is to download the entire message from the network, and extract the desired data from it. However, downloading the entire message, when it eventually stores only a fraction $\frac{1}{k}$ of it, is clearly wasteful of the network resources.

'Regenerating codes' [1] are a class of codes that aim to reduce the amount of download during repair, while retaining the storage efficiency of traditional MDS codes. Under the operation of a regenerating code, a replacement node connects to *any* $d$ ($\geq k$) existing nodes (termed *helper nodes*), and downloads $\beta$ symbols from each. This setting is illustrated in Fig. 1. With regenerating codes, the total amount of data $d\beta$ downloaded for repair is much smaller than the total size of the message $B$. It is shown in [1] that the parameters associated with a regenerating code must necessarily satisfy

$$B \leq \sum_{i=0}^{k-1} \min\left(\alpha, (d-i)\beta\right) \ . \quad (1)$$

A regenerating code is said to be *optimal* if it satisfies this bound with equality. Since both storage and bandwidth come at a cost, it is naturally desirable to minimize both $\alpha$ as well as $\beta$. However, it can be deduced (see [1]) that achieving equality in (1), for fixed values of $B$ and $[n, k, d]$, leads to a tradeoff between the storage space $\alpha$ and the amount of download for repair $d\beta$. The two extreme points in this tradeoff are termed the minimum storage regenerating (MSR) and minimum bandwidth regenerating (MBR) points. These points have been well studied in the literature, and several explicit constructions of codes operating at these points are available [2]–[8]. It has also been shown in [8] that essentially all other points on the tradeoff curve are not achievable.

In this paper, we address the problem of handling errors and erasures in distributed storage networks using regenerating codes. In particular, we are interested in codes that can perform reconstruction and efficient repair in the presence of errors and erasures at the nodes or in the links. Such codes are clearly useful in handling errors and packet losses occurring in the network. In addition, such codes can also be used to provide security in distributed storage systems, where malicious adversaries may corrupt the data stored in some nodes in the system.

K. V. Rashmi, Nihar B. Shah and Kannan Ramchandran are with the Dept. of EECS, University of California, Berkeley, CA 94703, USA. Email: {rashmikv, nihar, kannanr}@eecs.berkeley.edu. P. Vijay Kumar is with the Dept. of ECE, Indian Institute Of Science, Bangalore, India. Email: vijay@ece.iisc.ernet.in. P. Vijay Kumar is also an adjunct faculty member of the Electrical Engineering Systems Department at the University of Southern California, Los Angeles, CA 90089-2565.

This project is supported in part by an AFOSR grant (FA9550-10- 1-0567).

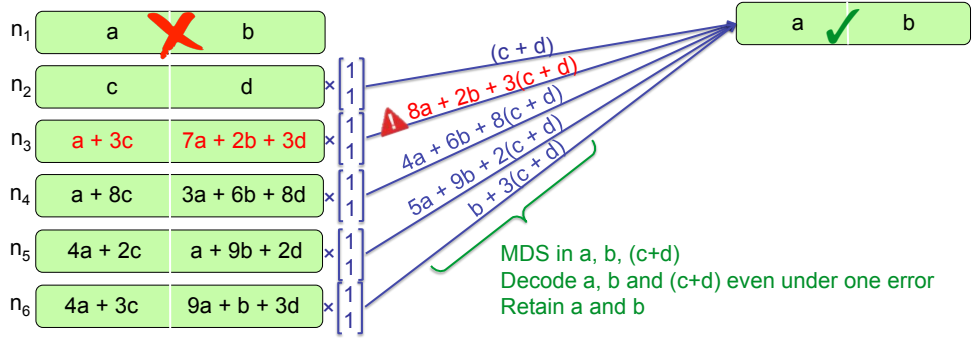

Fig. 2: An example of a universally resilient MSR code with parameters $[n=6, k=2, d=3]$, $(B=4, \alpha=2, \beta=1)$. Also depicted is an instance of one error correction during repair of node 1, by connecting to $(d+2)=5$ nodes.

The aspect of security in distributed storage systems employing regenerating codes is studied in [9], where an outer bound is provided for the total amount of data that can be stored securely in the presence of malicious adversaries. The model presented in [9] considers correction of a fixed number of errors, by designing encoding and storage algorithms specifically for this purpose. It is also shown that the code in [8] achieves this bound with an appropriate choice of the underlying MDS code, for the case $d = n-1$ at the MBR point. However, apart from this case, no other constructions of secure regenerating codes are known in the literature.

In the present paper, we provide a new approach for handling errors and erasures in regenerating codes. Under our system model, the data is encoded and stored assuming *no* error/erasure-resiliency requirements. The task of correcting the errors or erasures is performed in the decoding stage by downloading a larger amount of data. In contrast to [9], our approach allows for choosing a different level of resiliency during each event of repair or reconstruction, depending on the prevalent network state. A second advantage of our approach is that it allows for introducing resilience in regenerating codes that were *not* designed for handling errors and erasures.

We present explicit code constructions for the parameters (i) MSR, all $[n, k, d \geq 2k-2]$ and (ii) MBR, all $[n, k, d]$. In addition, we show the optimality of these codes through tight outer bounds on the storage and bandwidth requirements. This establishes the capacity of such systems for these parameters. Moreover, this also establishes the capacity of regenerating codes in the presence of malicious adversaries for these parameters, which had remained open. The decoding algorithms have a (polynomial) complexity, identical to that of Reed-Solomon codes. The codes presented here are based on a 'Product-Matrix' construction introduced in [2], that also possess other appealing properties such as linearity, scalability, and ease of implementation. An example of an MSR error/erasure resilient code is depicted in Fig. 2.

A natural question that follows is whether any regenerating code can be made resilient to errors and erasures in this fashion. In this paper, we also answer this question by providing necessary and sufficient conditions for a regenerating code to be resilient to errors and erasures. It turns out that, to date, the product-matrix codes are the only codes that satisfy these properties.

While we were writing this paper, we came across a contemporaneous independent work [10] that is related to the present paper, and deals with byzantine fault tolerance using product-matrix codes of [2]. The authors use a CRC to check the integrity of data during repair and reconstruction, and a feedback scheme to iteratively correct them. However, CRC based schemes are not applicable in settings such as protection against malicious adversaries, since the CRC can also be corrupted by the adversary. The present paper takes a more fundamental look at the problem of handling errors and erasures in regenerating codes.

The rest of the paper is organized as follows. The system model is described in Section II, and outer bounds for this model are also provided in this section. Explicit constructions of error-resilient regenerating codes are provided in Section III. Necessary and sufficient conditions for providing error and erasure resiliency in any regenerating code are presented in Section IV.

## II. SYSTEM MODEL

We consider a block-based model where the message is divided into blocks, and there is no coding across the blocks. All operations of encoding, decoding and repair are performed independently across the blocks. Thus the regenerating code parameters (for the error-free case) described in Section I can be considered as pertaining to a single block of data. More concretely, we consider a block to consist of $B$ message symbols, and the storage capacity in each of the $n$ nodes to be $\alpha$ symbols per block. One can reconstruct the $B$-message symbols by downloading the data pertaining to this block from any subset of $k$ nodes, and regenerate the data stored in any node by downloading $\beta$ symbols each (pertaining to this block) from any $d$ nodes.

We assume that the granularity of any error or erasure is one block. In other words, we assume that during repair, all the $\beta$ symbols passed by a helper node suffer the same fate: either all these $\beta$ symbols are erased, or all are in error, or all are perfectly received. Similarly, during reconstruction, all the $\alpha$ symbols passed by a node are assumed to suffer the same fate.

The assumption of block errors/erasures can accurately model several scenarios of interest, two of which are described here. Consider handling of packet drops during transmission across a network. The size of a packet during reconstruction and repair can be assumed to be a multiple of $\alpha$ and $\beta$ (since the packet size will usually be much larger than these parameters). Thus, when a packet is delayed or dropped, all the $\alpha$ or $\beta$ symbols corresponding to a block are erased. In the security scenario, to account for compromise of any node or link to a malicious adversary, one needs to protect against corruption of the entire data on that node or link. Thus, we can model the security scenario in our framework by simply considering the entire data as a single block.

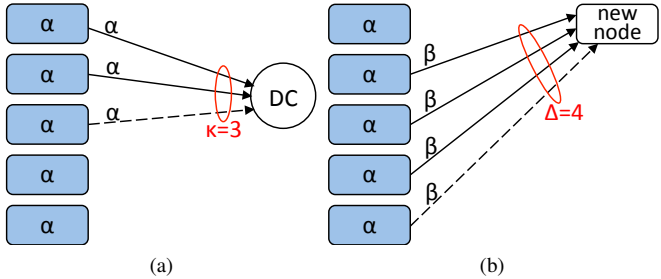

(a) (b)

Fig. 3: The system setting for an $(s = 1, t = 0)$-resilient regenerating code with $k = 2$ and $d = 3$. Here, connectivity during (a) reconstruction is $\kappa = k+1 = 3$, and (b) repair is $\Delta = d+1 = 3$.

We note that the absence of this assumption will allow arbitrarily scattered errors and erasures in the model. It can be shown that codes attempting to guard against such scattered errors/erasures require significantly larger overheads. As discussed above, many applications can be modelled as having block errors, thus avoiding these overheads.

We now define formally, the error/erasure handling capability of a regenerating code.

*Definition 1 ($(s, t)$-resilient code):* A regenerating code is $(s, t)$-resilient if it can correct upto $s$ erasures and $t$ errors during repair as well as reconstruction.

As discussed previously, under our system model, errors and erasures are corrected by downloading additional data during reconstruction or repair. One way to obtain additional data is to connect to a larger number of nodes, and we choose this approach. More precisely, we allow a connectivity of $\Delta (\geq d)$ nodes during repair and $\kappa (\geq k)$ nodes during reconstruction. The parameters $\Delta$ and $\kappa$ depend on the error/erasure correcting capability expected out of the system. Fig. 3 depicts the $(1, 0)$-resilient version of the system in Fig. 1.

We now provide an outer bound on the capacity of resilient regenerating codes. The bound is obtained by adapting the outer bound of [9] (for the omniscient adversary case) to our system model, and extending it to handle erasures as well.

*Theorem 1:* A $(s, t)$-resilient regenerating code, connecting to $\Delta$ and $\kappa$ nodes for repair and reconstruction respectively, must satisfy

$$B \leq \sum_{i=0}^{k-1} \min\left(\alpha, (d-i)\beta\right) \quad (2)$$

where $d = \Delta - s - 2t$ and $k = \kappa - s - 2t$.

*Proof (sketch):* The bound can be derived either using cut-set arguments in an information flow graph as in [1], [9] or using information theoretic arguments as in [8]. The complete proof is available in [11]. ∎

We will call an $(s, t)$-resilient regenerating code as *optimal* if it meets the bound in Theorem 1. Clearly, for $(s, t)$-resilience, we need

$$d + s + 2t \leq n - 1 \quad (3a)$$
$$k + s + 2t \leq n \quad (3b)$$

where (3a) represents the connectivity required during repair, and (3b) during reconstruction.

In many applications of interest, it may be desired to provide different levels of reliability during different instances of reconstruction and repair. For instance, under changing network states in the packet erasure setting, or varying threat levels under the security setting. We define codes that possess such a property as universally resilient codes.

*Definition 2 (Universally resilient code):* A regenerating code is universally resilient if it is simultaneously $(s, t)$-resilient for all $s$ and $t$ satisfying (3).

Thus a universally resilient code can correct upto $s$ erasures and $t$ errors by downloading $\beta$ (or $\alpha$) symbols each from $s+2t$ additional nodes during repair (or reconstruction), as long as $d + s + 2t \leq n - 1$ (or $k + s + 2t \leq n$).

In the next section, we present constructions of product-matrix codes that are universally resilient and optimal. Before moving on to the constructions, we briefly digress to explore some connections with network coding.

*Relation to Network Coding:* The regenerating codes problem described above, if relaxed to the requirement of repair of only the systematic nodes, turns out to be a non-multicast network coding problem (see [12, Section I-C]). While the occurrence of errors and erasures in multicast network coding are well studied in the literature [13]–[15], the results of the present paper lead to a class of non-multicast networks for which the error/erasure capacity of the network can be achieved by codes that are *linear*, deterministic and explicit.

III. ERROR/ERASURE-RESILIENT REGENERATING CODES

We provide explicit constructions of universally resilient MSR and MBR codes for
1) MSR, all parameters $[n, k, d \geq 2k - 2]$, and
2) MBR, all parameters $[n, k, d]$,

which meet the outer bound provided in Theorem 1. Thus, this also establishes the capacity of such a system for these parameter values. These codes are based on product-matrix (PM) codes that were introduced in [2].

As discussed in Section II, in our approach, the encoding algorithm is identical to the error/erasure free case. Hence, we first briefly describe the product-matrix code construction for the error/erasure free case [2], which meets the bound in (1). We then present the decoding algorithms (for both repair and reconstruction) that can handle $s$ erasures and $t$ errors for all values of $s$ and $t$ satisfying (3), with $\Delta = (d + s + 2t)$ and $\kappa = (k + s + 2t)$. These parameters satisfy the bound in Theorem 1 with equality, thereby establishing the optimality of these codes.

We begin with the minimum storage case, and subsequently present the minimum bandwidth case. The example depicted in Fig. 2 is an optimal, universally resilient MSR code.

## A. Universally resilient MSR Codes

MSR codes use the minimum possible storage at each node. Since the data from any $k$ nodes should suffice to reconstruct all the $B$ message symbols, each node must necessarily store at-least a fraction $\frac{1}{k}$ of the entire data. Hence for an MSR code we have $\alpha = \frac{B}{k}$. To meet the bound (1) with equality (in absence of errors/erasures), an MSR code must satisfy

$$B = k\alpha, \quad d\beta = \alpha + (k-1)\beta . \qquad (4)$$

In this section we present explicit constructions of optimal, universally resilient MSR codes for all parameter values $[n, k, d \geq 2k-2]$. The code is designed for the case $d = 2k-2$, which can be extended to any $d > 2k-2$ via the shortening technique for MSR codes provided in [2], [4]. When $d = 2k-2$, from (4), we get

$$\alpha = (k-1)\beta, \quad B = k(k-1)\beta . \qquad (5)$$

Since both $\alpha$ and $B$ are multiples of $\beta$, we obtain the optimal code for the desired parameters $(B, \alpha, \beta)$ by first constructing an optimal code for

$$\alpha' = (k-1), \quad B' = k(k-1) = \alpha'(\alpha'+1), \quad \beta' = 1 , \qquad (6)$$

and then concatenating this code $\beta$ times in parallel.

The PM-MSR code in [2] can be described in terms of an $(n \times \alpha')$ *code* matrix $C = \Psi M$, with the $i^{th}$ row of $C$ containing the $\alpha'$ symbols stored in node $i$. The $(n \times d)$ *encoding* matrix $\Psi$ is of the form $\Psi = [\Phi \ \Lambda\Phi]$, where $\Phi$ is an $(n \times \alpha')$ matrix and $\Lambda$ is an $(n \times n)$ diagonal matrix satisfying: (a) any $\alpha'$ rows of $\Phi$ are linearly independent, (b) any $d$ rows of $\Psi$ are linearly independent, and (c) the diagonal elements of $\Lambda$ are all distinct. The choice of the matrix $\Psi$ governs the choice of the finite field $\mathbb{F}_q$, e.g., choosing $\Psi$ as Vandermonde (carefully chosen to satisfy condition (c)) permits any $q \geq 4n$. The $((d = 2\alpha') \times \alpha')$ *message* matrix $M$ is of the form $M = [S_1 \ S_2]^t$, where $S_1$ and $S_2$ are $(\alpha' \times \alpha')$ symmetric matrices. The superscript $t$ is used to denote the transpose of a vector or matrix. The two symmetric matrices $S_1$ and $S_2$ together contain $\alpha'(\alpha'+1)$ distinct elements, which are populated by the $B = \alpha'(\alpha'+1)$ message symbols. This completes the description of the encoding algorithm.[1]

The following theorems show that this code is optimally universally resilient during repair and reconstruction.

*Theorem 1 (MSR Repair):* In the MSR code presented, the $\alpha$ symbols stored in any node can be recovered by downloading $\beta$ symbols each from *any* $\Delta = d + s + 2t$ nodes, in the presence of upto $s$ (block) erasures and $t$ (block) errors.

*Proof:* Since we consider only block errors and erasures, it suffices to describe the repair algorithm for the code with $\beta' = 1$, and the same algorithm is applied in parallel to obtain the repair algorithm for the desired code. Consider failure of node $f$ in the system, and let $\begin{bmatrix} \underline{\phi}_f^t & \lambda_f \underline{\phi}_f^t \end{bmatrix}$ be the row of $\Psi$ corresponding to the failed node. Thus the $\alpha'$ symbols stored in node $f$ are

$$\begin{bmatrix} \underline{\phi}_f^t & \lambda_f \underline{\phi}_f^t \end{bmatrix} M = \underline{\phi}_f^t S_1 + \lambda_f \underline{\phi}_f^t S_2 . \qquad (7)$$

[1]The code can be converted to a systematic form (such as in Fig. 2) by a simple symbol remapping technique as shown in [2, Section V-B].

The replacement for the failed node $f$ connects to an arbitrary set $\{h_j \mid j = 1, \ldots, \Delta\}$ of $\Delta$ nodes. To facilitate repair of node $f$, node $h_j$ computes the inner product $\underline{\psi}_{h_j}^t M \underline{\phi}_f$ and passes on this value to the replacement node. Letting $\underline{m}_f = M\underline{\phi}_f$, we can write the symbol passed by node $h_j$ as $\underline{\psi}_{h_j}^t \underline{m}_f$. Thus the $\Delta$ symbols obtained at the destination are $\Psi_{\text{rep}} \underline{m}_f$, where

$$\Psi_{\text{rep}} = \begin{bmatrix} \underline{\psi}_{h_1} & \underline{\psi}_{h_2} & \cdots & \underline{\psi}_{h_\Delta} \end{bmatrix}^t .$$

Since any $d$ rows of $\Psi$ are linearly independent by construction, and since $\Psi_{\text{rep}}$ comprises a subset of the rows of $\Psi$, $\Psi_{\text{rep}} \underline{m}_f$ is simply an MDS encoding of the $d$ symbols in the vector $\underline{m}_f$. It follows that this code has a minimum distance of $(\Delta - d + 1) = (s + 2t + 1)$ which allows us to recover $\underline{m}_f$ using standard decoding algorithms [16] in the presence of upto $s$ erasures and $t$ errors. Thus the replacement node now has access to

$$\underline{m}_f = M\underline{\phi}_f = \begin{bmatrix} S_1 \underline{\phi}_f \\ S_2 \underline{\phi}_f \end{bmatrix} .$$

Since $S_1$ and $S_2$ are symmetric matrices, the replacement node has access to $\underline{\phi}_f^t S_1$ and $\underline{\phi}_f^t S_2$. Using this it can obtain $\underline{\phi}_f^t S_1 + \lambda_f \underline{\phi}_f^t S_2$, which is precisely the data previously stored in node $f$. ∎

*Theorem 2 (MSR Reconstruction):* In the MSR code presented, a data-collector can reconstruct all the $B$ message symbols by downloading data stored in any $\kappa = k + s + 2t$ nodes in the presence of upto $s$ (block) erasures and $t$ (block) errors.

*Proof (sketch):* The data reconstruction property of the code in the error-free case, as shown in [2], implies that the data passed by the $\kappa$ nodes are MDS over the finite field $\mathbb{F}_q^\alpha$. Over this finite field, the message is of size $k$, and the minimum distance of this MDS code is $(\kappa - k + 1) = (s + 2t + 1)$. This guarantees reconstruction of the $k$ source symbols over $\mathbb{F}_q^\alpha$, and equivalently the $k\alpha = B$ source symbols over $\mathbb{F}_q$, in the presence of upto $s$ erasures and $t$ errors. ∎

Explicit data-reconstruction algorithms are provided in [11].

## B. Universally resilient MBR Codes

MBR codes achieve minimum possible download during repair: a replacement node downloads only what it stores, resulting in $d\beta = \alpha$. To meet the bound (1) with equality (in absence of errors/erasures) an MBR code must satisfy

$$B = \left(kd - \binom{k}{2}\right)\beta, \ \alpha = d\beta . \qquad (8)$$

In this section we present explicit constructions of optimal, universally resilient MBR codes for all parameter values $[n, k, d]$. As in the MSR case, $B$ and $\alpha$ are multiples of $\beta$, and we first construct codes for

$$B' = \left(kd - \binom{k}{2}\right), \quad \alpha' = d, \quad \beta' = 1 . \qquad (9)$$

The desired code can be obtained by concatenating $\beta$ copies of this code.

The PM-MBR code in [2] has a similar form, $C = \Psi M$, as the PM-MSR code. The MBR code has the $(n \times d)$ encoding matrix $\Psi$ of the form $\Psi = [\Phi \ \Sigma]$, where $\Phi$ is an $(n \times k)$ matrix satisfying: (a) any $k$ rows of $\Phi$ are linearly independent, (b) any $d$ rows of $\Psi$ are linearly independent. For instance, one can choose $\Psi$ to be a Vandermonde matrix. The $(d \times d)$ message matrix $M$ is symmetric and consists of the $B'$ message symbols arranged as

$$M = \begin{bmatrix} S & T \\ T^t & 0 \end{bmatrix}.$$

Here, the $((d-k) \times k)$ matrix $T$ and the $(k \times k)$ *symmetric* matrix $S$ contain the $B' = kd - \binom{k}{2} = k(d-k) + \frac{k(k+1)}{2}$ message symbols as their elements.

The following theorems show that this code is optimally universally resilient during repair and reconstruction.

*Theorem 3 (MBR Repair):* In the MBR code presented, the $\alpha$ symbols stored in any node can be recovered by downloading $\beta$ symbols each from *any* $\Delta = (d + s + 2t)$ nodes, in the presence of upto $s$ (block) erasures and $t$ (block) errors.

  *Proof:* As in the case of MSR, it is sufficient to describe the repair algorithm for the code with $\beta' = 1$. Consider failure of node $f$ in the system, and let $\underline{\psi}_f^t$ be the row of $\Psi$ corresponding to the failed node. Thus the $\alpha'$ symbols stored in node $f$ are $\underline{\psi}_f^t M$. We will follow the notation as in Theorem 1. The helper node $h_j$ passes the symbol $\underline{\psi}_{h_j}^t M \underline{\psi}_f$. Denoting $\underline{m}_f = M \underline{\psi}_f$, the $\Delta$ symbols obtained at the destination can be written as $\Psi_{\text{rep}} \underline{m}_f$ where

$$\Psi_{\text{rep}} = \begin{bmatrix} \underline{\psi}_{h_1} & \underline{\psi}_{h_2} & \ldots & \underline{\psi}_{h_\Delta} \end{bmatrix}^t.$$

By construction, $\Psi_{\text{rep}} \underline{m}_f$ corresponds to an MDS encoding of the vector $\underline{m}_f$. As in the case of MSR, this code has minimum distance of $(\Delta - d + 1) = (s + 2t + 1)$ which allows us to recover $\underline{m}_f$ in the presence of upto $s$ erasures and $t$ errors. Since the message matrix $M$ is symmetric, $\underline{m}_f^t = \underline{\psi}_f^t M^t = \underline{\psi}_f^t M$ is precisely the set of $\alpha'$ symbols required. ∎

*Theorem 4 (MBR Reconstruction):* In the MBR code presented, a data-collector can reconstruct all the $B$ message symbols by downloading data stored in any $\kappa = (k + s + 2t)$ nodes in the presence of upto $s$ (block) erasures and $t$ (block) errors.

  *Proof (sketch):* As in the MSR case, the proof exploits the reconstruction property of PM-MBR codes in the error-free case [2]. The reconstruction property implies that, over $\mathbb{F}_q^\alpha$, the minimum distance of the code is $(\kappa - k + 1) = (s + 2t + 1)$. This guarantees reconstruction of the $k$ symbols over $\mathbb{F}_q^\alpha$, and equivalently the $B$ source symbols over $\mathbb{F}_q$, in the presence of upto $s$ erasures and $t$ errors. ∎

Explicit data-reconstruction algorithms are provided in [11].

## IV. NECESSARY AND SUFFICIENT CONDITIONS

The conversion of the product-matrix codes into universally resilient codes, as described in Section III, raises a natural question as to whether *any* regenerating code can be made universally resilient in a similar manner. We answer this question by providing a necessary and sufficient condition for the same. To the best of our knowledge, the only codes today that satisfy this condition are the product-matrix codes.

*Theorem 5:* An $[n, k, d]$ regenerating code can be made universally resilient if and only if the following condition holds: during any instance of repair, the data passed by a node $h$ helping in the repair, to the failed node $f$, depends only on $h$ and $f$, and not on the identities of the other nodes helping in this repair.

The proof of the theorem is provided in [11].

*Remark 1:* Clearly, since the number of nodes contacted during repair must satisfy $(d + s + 2t) \leq (n - 1)$, the requirement of having either $s > 0$ or $t > 0$ requires that $n > (d+1)$. Thus the code should not restrict the number of nodes $n$ to be $(d-1)$.

The only explicit regenerating codes that support $n > (d+1)$ are the high-rate 'approximately-exact' MSR codes of [3] and the product-matrix codes [2]. However, the MSR codes of [3] do not satisfy the condition provided in Theorem 5. As shown in Section III, the product-matrix codes satisfy this condition.

ACKNOWLEDGEMENT

The authors would like to thank Salim El Rouayheb and Sameer Pawar for fruitful discussions.